\documentclass[preprint2]{emulateapj}
\usepackage{multirow}
\usepackage{amsmath}

\usepackage{graphicx}
\newcommand{\Reff}{$R_{e}$}

\newcommand{\kms}{km~s$^{-1}$}
\newcommand{\magarcsec}{mag~arcsec$^{-2}$}

\newcommand{\vlsbb}{VLSB$-$B}
\newcommand{\vlsbd}{VLSB$-$D}

\shorttitle{Dark Matter in Virgo UDGs}
\shortauthors{Toloba et al.}

\begin{document}

\title{Dark Matter in Ultra Diffuse Galaxies in the Virgo Cluster from
their Globular Cluster Populations}


\author{Elisa~Toloba\altaffilmark{1}}\email{etoloba@pacific.edu}
\author{Sungsoon~Lim\altaffilmark{2,3}}
\author{Eric~Peng\altaffilmark{2,3}}
\author{Laura~V.~Sales\altaffilmark{4}}
\author{Puragra~Guhathakurta\altaffilmark{5}}
\author{J.~Christopher~Mihos\altaffilmark{6}}
\author{Patrick~C$\rm{\hat{o}}$t\'e\altaffilmark{7}}
\author{Alessandro~Boselli\altaffilmark{8}}
\author{Jean-Charles~Cuillandre\altaffilmark{9}}
\author{Laura~Ferrarese\altaffilmark{7}}
\author{Stephen~Gwyn\altaffilmark{7}}
\author{Ariane~Lan\c{c}on\altaffilmark{10}}
\author{Roberto~Mu\~noz\altaffilmark{11}}
\author{Thomas~Puzia\altaffilmark{11}}

\affil{$^1$Department of Physics, University of the Pacific, 3601
  Pacific Avenue, Stockton, CA 95211, USA}
\affil{$^2$Department of Astronomy, Peking University, Beijing 100871,
  China}
\affil{$^3$Kavli Institute for Astronomy and Astrophysics, Peking
  University, Beijing 100871, China}
\affil{$^4$Department of Physics and Astronomy, 900 University Avenue,
Riverside, CA 92521, USA}
\affil{$^5$UCO/Lick Observatory, University of California, Santa Cruz, 1156 High Street, Santa Cruz, CA 95064, USA}
\affil{$^6$Department of Astronomy, Case Western Reserve University,
  Cleveland, OH 44106, USA}
\affil{$^7$National Research Council of Canada, Herzberg Astronomy and
  Astrophysics Program, Victoria, BC V9E 2E7, Canada}
\affil{$^8$Aix Marseille University, CNRS, LAM, Laboratoire d’Astrophysique de Marseille, Marseille, France}
\affil{$^9$CEA/IRFU/SAp, Laboratoire AIM Paris-Saclay, CNRS/INSU, Universit\'e Paris Diderot, Observatoire de Paris, PSL Research
University, F-91191 Gif-sur-Yvette Cedex, France}
\affil{$^{10}$Observatoire Astronomique de Strasbourg, Universit\'e de Strasbourg, CNRS, UMR 7550, 11 rue de l'Universit\'e, F-67000 Strasbourg,
France}
\affil{$^{11}$Institute of Astrophysics, Pontificia Universidad Cat\'olica de Chile, Av. Vicu\~na Mackenna 4860, 7820436 Macul, Santiago, Chile}

\begin{abstract}

We present Keck/DEIMOS spectroscopy of globular clusters (GCs) around
the ultra-diffuse galaxies (UDGs) \vlsbb, \vlsbd, and VCC615 located
in the central regions of the Virgo cluster. We spectroscopically identify 4, 12, and 7 GC
satellites of these UDGs, respectively. We find that the three UDGs
have systemic velocities ($V_{sys}$) consistent with being in the Virgo cluster,
and that they span a wide range of velocity dispersions, from $\sim
16$ to $\sim 47$~\kms, and high dynamical mass-to-light ratios within the
radius that contains half the number of GCs ($ 407^{+916}_{-407}$,
$21^{+15}_{-11}$, $60^{+65}_{-38}$, respectively).
\vlsbd\ shows possible evidence for rotation along the stellar major axis and its
$V_{sys}$ is consistent with that of the massive galaxy M84 and
the center of the Virgo cluster itself. These
findings, in addition to having a dynamically and spatially ($\sim 1$~kpc)
off-centered nucleus and being extremely elongated, suggest that
\vlsbd\ could be tidally perturbed. 
On the contrary, \vlsbb\ and VCC615 show no signals of tidal deformation. Whereas the dynamics of \vlsbd\ suggest that it has a
less massive dark matter halo than expected for its stellar mass,
\vlsbb\ and VCC615 are consistent with a $\sim 10^{12}$~M$_{\odot}$ dark matter
halo. Although our samples of galaxies and GCs are small, these results suggest that UDGs may be a diverse population,
with their low surface brightnesses being the result of very early
formation, tidal disruption, or a combination of the two.

\end{abstract}


\section{Introduction}

Ultra diffuse galaxies (UDGs) are extremely low surface brightness
galaxies (central surface brightness $\mu_{g,0}\gtrsim24$~\magarcsec)
with luminosities in the dwarf galaxies regime ($M_V\gtrsim-16$), and sizes in the massive
galaxies regime (half-light radius \Reff~$\gtrsim 1.5$~kpc). UDGs
are characterized by spheroidal shapes, nearly exponential surface brightness
profiles, and quenched stellar populations \citep{vD15a,Mihos15,Mihos17}. 

Large low surface brightness galaxies were found for the
first time in the Virgo cluster photographic plates by
\citet{SandageBinggeli84,Binggeli87}. Later on, other studies found a few more of
these diffuse galaxies
\citep{Impey88,Dalcanton97,Caldwell06}. However, with
the new deep imaging surveys, a plethora of these systems are
being found mainly in cluster environments \citep{Koda15,Mihos15,Munoz15,vD15a,MartinezDelgado16,Toloba16b,vanderburg16, Janssens17,Mihos17,RomanTrujillo17,venhola17}.

There are three main posible mechanisms that could explain the
observed properties of the UDGs. (1) They could be extended dwarf
galaxies. Some simulations predict them to be rapidly rotating
\citep{AmoriscoLoeb16}, while others suggest that their extended sizes
are the result of strong gas outflows \citep{DiCintio17}.
(2) They could be tidal galaxies formed from the debris of
harassed and ram pressure stripped galaxies that lost large
fractions of stars. In these two scenarios, the UDGs are expected to have
shallow potential wells which makes them vulnerable to the
cluster environment \citep[e.g.,][]{Moore96}. However, UDGs are found in extremely dense
regions such as the core of the Virgo and Fornax clusters \citep{Mihos15,Mihos17,Munoz15} and the Coma
cluster \citep{vD15a}. They could be falling in the
cluster for the first time. (3) They are ``failed'' massive galaxies where the environment
and/or internal feedback stopped the star formation and, as a result,
the number of stars is smaller than expected for that
size. In this scenario, UDGs have a massive dark matter halo
that makes them less prone to disruption.

The large number of globular clusters (GCs) found in some UDGs
\citep{vD17} points to the third scenario given that these numbers are
more typical of massive galaxies than dwarfs \citep{Peng06}. However,
an analysis of a larger sample of UDGs suggests that they do not have a
statistically significant excess of GCs compared to normal dwarf
galaxies of the same stellar mass \citep{Amorisco17}.

Measuring the dark matter halo would help to distinguish between
formation scenarios.
A massive dark halo can explain
their survival in high density environments and their origin as
``failed'' massive galaxies gets stronger. A low mass dark halo
would suggest that UDGs are puffed up
dwarf galaxies that are likely on the verge of disruption. However, if
disruption is currently happening, it is hard to interpret dark matter
halo mass estimates based on observed velocities.

There are three UDGs with kinematic measurements in the
literature. All three seem to have
massive dark matter halos
\citep[$10^{10}-10^{11}$~M$_{\odot}$,][]{Beasley16,vD16,vD17}. We
analyze here the internal dynamics of three UDGs in the central regions of
the Virgo cluster doubling the current statistics. We assume the
distance to the Virgo cluster is 16.5~Mpc \citep{Mei07,Blakeslee09}.

\section{Data}

\subsection{Sample Selection}

We target GC candidates in the Virgo UDGs \vlsbb, \vlsbd,
and VCC615 \citep[][]{Binggeli87,Mihos15,Mihos17}. The GCs are selected from the Next Generation Virgo cluster Survey
\citep[NGVS;][]{Ferrarese12}. Point-like sources are split
into three categories attending to their probability of being
foreground stars, GCs in the Virgo cluster, and background
galaxies. These probabilities are obtained combining the position
  of all point-like sources in different color-color diagrams based on
$u*$, $g$, $i$, $z$ photometry \citep[and $Ks$ only available for \vlsbb,][]{Munoz14} with the
inverse concentration parameter (ic) which measures how point-like or
extended the object is \citep[see Figure~\ref{images};][Peng et~at., in
prep]{Powalka16}.

We select objects with $g<24.5$ and with higher probability of being
GCs than being foreground stars or background galaxies. Due to the
large field-of-view of the DEIMOS spectrograph ($16.3' \times 5'$), we
also include some foreground stars that, due to their
position on the sky, are candidates for Virgo Overdensity and
Sagittarius Streams (Figure~\ref{images}). Their analysis will be presented in a future paper.

\subsection{Observations and Data Reduction}

The observations were carried out with the DEIMOS spectrograph
\citep{Faber03} located at the KeckII 10m telescope (Mauna Kea
Observatory). We designed one mask per UDG and the
600~lines/mm grating centered at 7200~\AA\ with slit widths of $1''$
and the GG455 blocking filter. The wavelength coverage is
$4700-9200$~\AA\ with a pixel scale of 0.52~\AA/pixel, and a spectral
resolution of 2.8~\AA\ (FWHM). 

The three DEIMOS slitmasks had position angles (P.A.) of 105~deg,
$-152$~deg, and $-140$~deg, respectively for \vlsbb, \vlsbd, and VCC615. All the slits were aligned with the
slitmasks but for VCC615, for which the slits had a 10~deg offset,
resulting in the slits having P.A.~$= -130$~deg.

The slitmasks were observed on March~04~2017, with exposure times of 83~min for \vlsbb, 78~min for \vlsbd, and
87~min for VCC615. The average seeing was $0.6''$ (FWHM).

We reduced the data with the {\sc spec2d} pipeline
\citep{DEIMOSpipeline1,DEIMOSpipeline2} with improvements described
by \citet{Kirby15a,Kirby15b}. The wavelength solution is improved by tracing the sky lines along the
slit and improving the extraction of the one-dimensional spectra by
accounting for the differential atmospheric refraction along the slit.
The main steps in the reduction process consisted of flat-field corrections, wavelength calibration, sky subtraction, and
cosmic ray cleaning.

\begin{table}
\begin{center}
\caption{Properties of the UDGs \label{table}}
{\renewcommand{\arraystretch}{1.}
\resizebox{8.5cm}{!} {
\begin{tabular}{l|c|c|c}
\hline \hline \\ 
         &  VLSB-B  &  VLSB-D  &  VCC615 \\ 
\hline \\ 
RA (hh:mm:ss) & 12:28:10.6 & 12:24:42.1 & 12:23:04.7 \\ 
DEC (dd:mm:ss) & $+$12:43:28 & $+$13:31:02 & 12:00:56 \\ 
M$_V$ (mag) & $-13.5 \pm 0.2$ & $-16.2\pm 0.4$ & $-14.7 \pm 0.1$ \\ 
$R_e$ (kpc) & $2.9\pm 0.2$ & $13.4 \pm 2.0$ & $2.4 \pm 0.1$ \\ 
$\epsilon$  & $0.17\pm  0.15 $ & $0.55\pm 0.10$ & $0.05\pm 0.05$ \\
$\langle \mu_{V} \rangle_e$ (mag~arcsec$^{-2}$) & $27.5 \pm 0.1$
                    & $27.6 \pm 0.2$ & $25.8 \pm 0.1$ \\ 
$M*$~($\times10^7$M$_{\odot}$) & $0.6\pm 0.1$ & $7.9\pm 0.1$ & $2.1\pm 0.1$ \\ 
$R_h$ (kpc) & $1.8^{+0.8}_{-0.6}$ & $8.4^{+8.7}_{-2.8}$ & $1.9^{+0.7}_{-0.5}$ \\ 
$N_{GC,tot}$ & $12^{+7}_{-5}$ & $36^{+47}_{-17}$ & $14^{+6}_{-5}$ \\ 
\hline \\ 
$N_{GC,spec}$ &  4  & 12  &  7  \\ 
$V$~(\kms) & $24.9^{+22.3}_{-36.2}$ & $1033.8^{+5.9}_{-5.5}$ & $2094.0^{+14.9}_{-13.0}$ \\ 
$V_{nuc}$~(\kms) & $-$ & $1040.1\pm1.4$ & $2094.1\pm 2.7$ \\ 
$\sigma$~(\kms) & $47^{+53}_{-29}$ & $16^{+6}_{-4}$ & $32^{+17}_{-10}$ \\ 
$\frac{dv}{dr}$~(\kms\ arcmin$^{-1}$) & $-$ & $5.9^{+11.7}_{-11.9}$ & $-$ \\ 
$V_{rot}$~(\kms) & $-$ & $17.2^{+33.9}_{-34.7}$ & $-$ \\ 
$M_{1/2}$~($\times10^9$M$_{\odot}$) & $4.9^{+11.1}_{-4.9}$ & $3.2^{+2.4}_{-1.7}$ & $2.5^{+2.7}_{-1.6}$ \\ 
$M/L_V$~(M$_{\odot}/$L$_{\odot}$) & $407^{+916}_{-407}$ & $21^{+15}_{-11}$ & $60^{+65}_{-38}$ \\ 
$f_{DM}$~($\%$) & $>99\pm1$ & $99\pm 1$ & $>99 \pm 1$\\ 
\hline \\
\end{tabular}
}}
\end{center}
\tablecomments{Rows $1-8$: photometric
  parameters. Rows $9-18$: spectroscopic measurements. The
  central coordinates are in
  J2000. The magnitudes are in the Vega system. $R_e$ is the stellar
  half-light radius. $\epsilon$ is the ellipticity. $\langle \mu_{V} \rangle_e$ is the average
  surface brightness within the $R_e$. These three parameters are
  taken from \citet{Mihos15,Mihos17}. $M*$ is the total stellar
  mass. $R_h$ is the radius that contains half the total number of
  GCs. $N_{GC,tot}$ is the total number of GCs. 
  $N_{GC.spec}$ is the number of spectroscopically confirmed GCs. The remaining candidates were not observed or had too low S/N
  to estimate reliable velocities. $V$ is the heliocentric systemic velocity. $V_{nuc}$ is the heliocentric velocity for the
  nucleus. $\sigma$ is the velocity dispersion of the GC system (it includes
  rotation if present). $\frac{dv}{dr}$ is the velocity gradient
  along the P.A. indicated in Figure~\ref{MCMC}. $V_{rot}$ is the
  rotation speed derived from the velocity gradient. $M_{1/2}$ is the dynamical
  mass within the $R_h$. $M/L_V$ is the dynamical mass-to-light ratio within the
 $R_h$. $f_{DM}$ is the dark matter fraction within the $R_h$ and its
 error bar refers to the $99\%$ confidence interval.} 
\end{table}

\section{Results}

\subsection{Radial Velocity Measurements and Membership Criteria} 

Line-of-sight radial velocities are measured following the
same steps as described in \citet{Toloba16}. In short, we feed the penalized
pixel-fitting software \citep[pPXF;][]{ppxf} with 17 high signal-to-noise
($100<S/N<800$~\AA$^{-1}$) stellar templates observed with the same instrumental setup as
the science data. To reduce the mismatch fitting problem, the stellar templates include stellar types from 
B1 to M0 and luminosity classes from supergiants to dwarfs.
The radial velocity uncertainties are calculated running 1000 Monte
Carlo simulations where the flux of each spectrum is perturbed within
the flux uncertainty obtained during the reduction process assuming
that it is Gaussian.

The final radial velocities are corrected by small offcentering
effects across the slits. This affects unresolved sources and is
quantified using the atmospheric B and A bands at $6850-7020$~\AA\ and
$7580-7690$~\AA. The resulting radial velocity uncertainties are the
quadrature sum of the uncertainty in the observed radial velocity and
the A and B bands.

\begin{figure*}[p]
\centering 
\includegraphics[angle=0,width=18cm,bb=39 99 575 629]{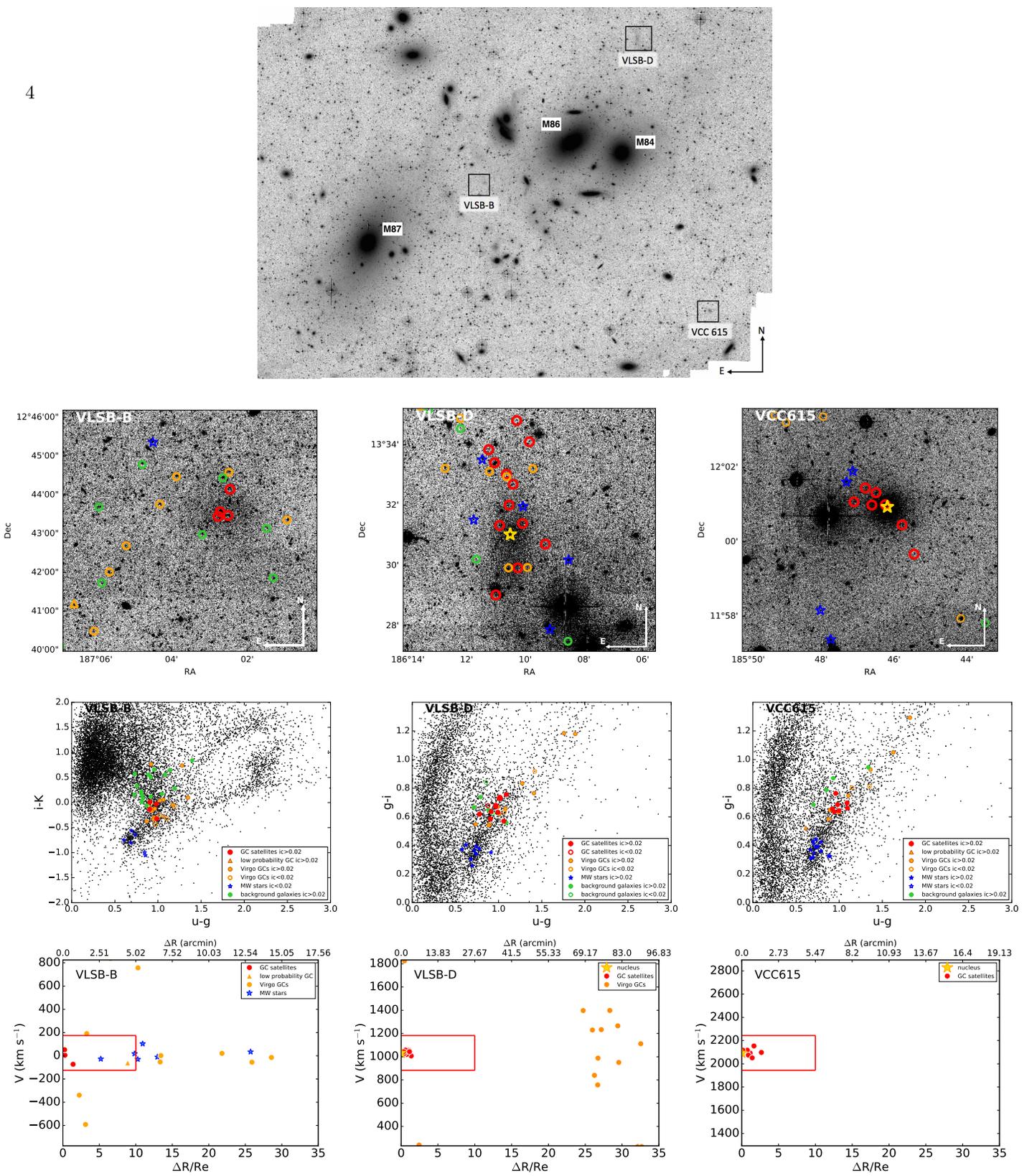}
\vspace{0.5cm}
\caption{Upper panel: g band image for the central $3\times2$~deg of
  the Virgo cluster from \citet{Mihos17}. Second row of panels: NGVS g band
  images of the three UDGs. Different colors indicate objects of
  different nature based on their spectrophotometric
  properties (symbols as in the lowest panels). 
Note
  that \vlsbb\  is not nucleated. Third row of panels:
    color-color diagrams. Symbols are split into high inverse
    concentration (ic; i.e. extended) and low ic (point-like)
    sources. The K band, only available for \vlsbb, clearly separates
    the GC and stellar locus. The blue ($u-g\sim 0.2$) vertical band
   is the locus of background galaxies. Lower panels: membership diagrams. 
The orange triangle in \vlsbb\ indicates an object
whose probability of being a GC based on multi-wavelength photometry,
that includes the $K$ band, and the extreme deconvolution technique is
just slightly higher than its probability of being a MW star ($60\%$
vs. $38\%$). To be cautious, we will not consider this object as a GC satellite.}\label{images}
\end{figure*}

The membership criteria is described in \citet{Toloba16}
and summarized in Figure~\ref{images}. Those GCs that are within a box
of $\Delta R/R_e<10$ and $\mid\Delta V\mid<150$~\kms, approximately three times the
typical velocity dispersion of dwarf galaxies, are considered GC satellites. The expected contamination for all
the GC candidates combined is $1.1\pm0.2$ within this box. The
contaminants would be intracluster GCs, GC satellites of other galaxies in Virgo, and Milky
Way stars. Background galaxies are spectroscopically identified for
their emission lines and removed from the sample. The
GCs classified as satellites have median photometric probabilities of being
GCs of $91\%$ and photometric probabilities of being stars smaller than $5\%$. 

\begin{figure*}[p]
\centering 
\includegraphics[angle=0,width=18cm]{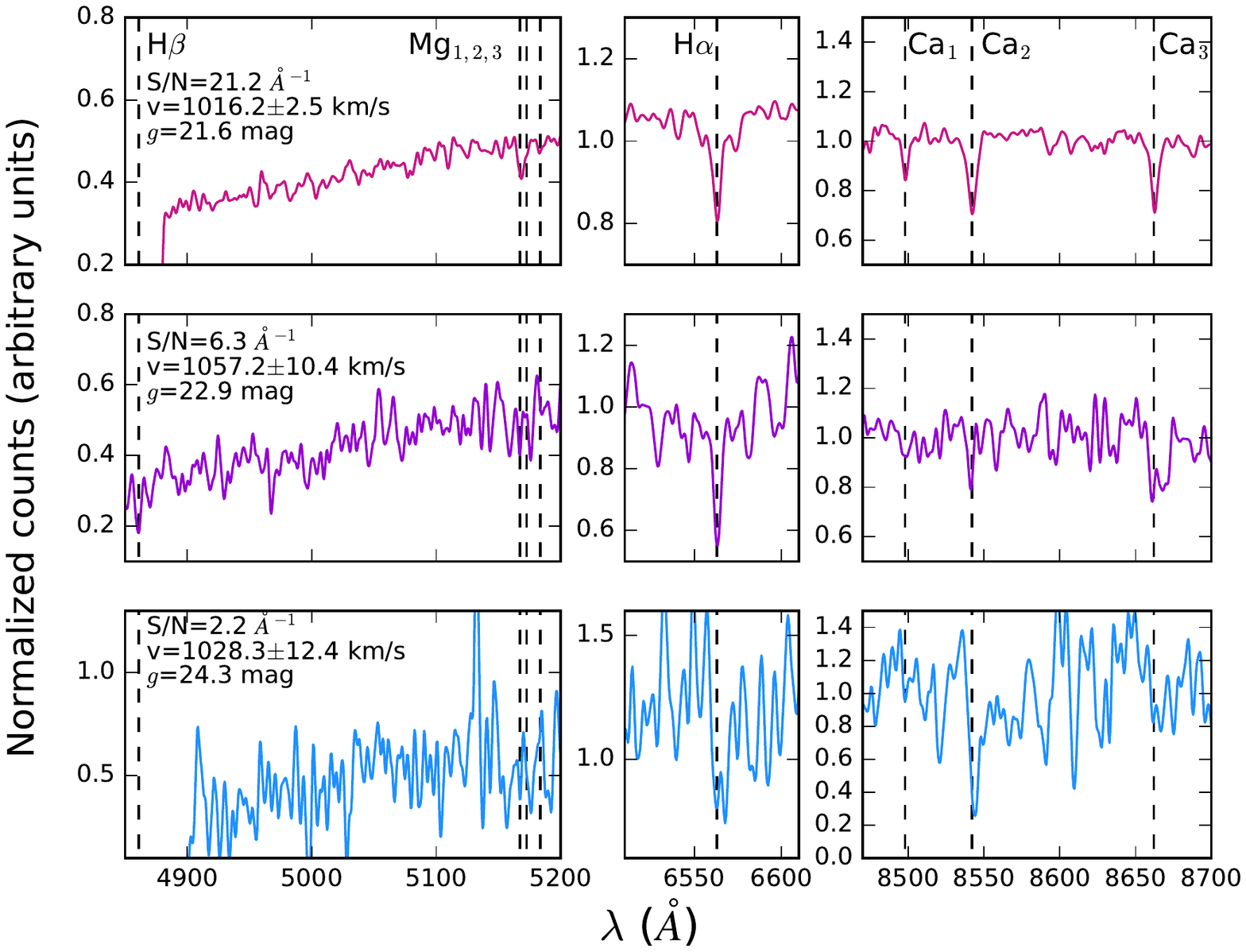}
\vspace{0.5cm}
\caption{Examples of three GC spectra with different S/N put into
    the rest frame using their radial velocities. The panels are
    arranged in order of decreasing S/N and luminosity from top to
    bottom. In each panel the S/N, heliocentric velocity, and $g$ band
    magnitude are shown. Three wavelength regions are shown for each spectrum: the region that includes the H$\beta$ and the Mg triplet lines, the region that includes the H$\alpha$ line, and the region that includes the Ca triplet lines. These lines are indicated with vertical dashed black lines.}\label{spectra}
\end{figure*}

\subsection{Velocity Dispersion and Velocity Gradient}

The dynamical properties of the three UDGs are analyzed using the Markov
Chain Monte Carlo (MCMC) method \citep{emcee}. We make two implementations to avoid
having more than two free parameters at a time. Assuming that the
line-of-sight radial velocities ($v$) come from a Gaussian distribution, the
logarithmic probability of the observed velocities for a certain systemic velocity
($V_{sys}$) and velocity dispersion ($\sigma$) is:

\begin{equation}\label{eqn1}
\mathcal{L} (V_{sys}, \sigma)=-\frac{1}{2}\sum_{n=1}^N \log (2\pi (\sigma^2+\delta
v_n^2)) - \sum_{n=1}^N \frac{(v_n-V_{sys})^2}{2(\sigma^2+\delta v_n^2)}
\end{equation}

\noindent where $N$ is the number of GC satellites and $\delta v$ are
the radial velocity uncertainties which contribute to increase the width of the
Gaussian distribution.

We run the implementation based on Equation~\ref{eqn1} twice. The
second time $V_{sys}$ is fixed to the heliocentric
velocity of the nucleus. This can only be done for VCC615 and \vlsbd,
the two nucleated UDGs. 

Given the low numbers of GC satellites, we perform simulations to
test the statistical significance and possible biases in the
calculations. We simulate Gaussian distributions with input dispersions from
10 to 100~\kms\ in steps of 10~\kms. For each one of these
distributions, we randomly select 4, 7, and 12 velocities with
uncertainties that are the average uncertainty of our observed radial
velocities. For each randomly selected sample, we apply the same MCMC
method as described above. In general, the input velocity dispersion
is always recovered with possibly a small bias of $\sim 5$~\kms\ for
large input velocity dispersions: $>80$~\kms\ for samples of 12 GC
satellites; $>40$~\kms\ for samples of 7 GC satellites. For samples of
4 GC satellites and input dispersions $>40$~\kms, the overestimation can be as
high as $10$~\kms. However, this small bias is always
within the measured error bars. 
On the contrary, if $V_{sys}$ is fixed and
$\sigma$ is the only free parameter, the difference between the input
and output dispersion is always $\leq 5$~\kms\ for input dispersions
$\leq 60$~\kms.

We estimate whether \vlsbd, the UDG with the largest number of
spectroscopically confirmed GC satellites, shows internal rotation using the MCMC implementation
described in \citet{Martin10}. The logarithmic probability in this
case is:

\begin{equation}\label{eqn2}
\begin{split}
\mathcal{L} (V_{sys}, \sigma, dv/dr, \phi) = & -\frac{1}{2}\sum_{n=1}^N \log (2\pi (\sigma^2+\delta
v_n^2)) \\ & - \sum_{n=1}^N
\frac{(v_n-V_{sys}-\frac{dv}{dr}r_n)^2}{2(\sigma^2+\delta v_n^2)}
\end{split}
\end{equation}

\noindent where $V_{sys}$ is fixed to the value obtained running the
MCMC in Equation~\ref{eqn1} and $dv/dr$ is the velocity gradient along the projected
distance $r$ with P.A.~$=\phi$:

\begin{equation}
r=(RA-RA_0)\cos (Dec_0) \sin (\phi) +(Dec-Dec_0)\cos (\phi)
\end{equation}

\noindent $RA_0$ and $Dec_0$ are the coordinated of the photometric galaxy center.

\begin{figure*}
\centering
\includegraphics[angle=0,width=18cm,bb=46 139 601 538]{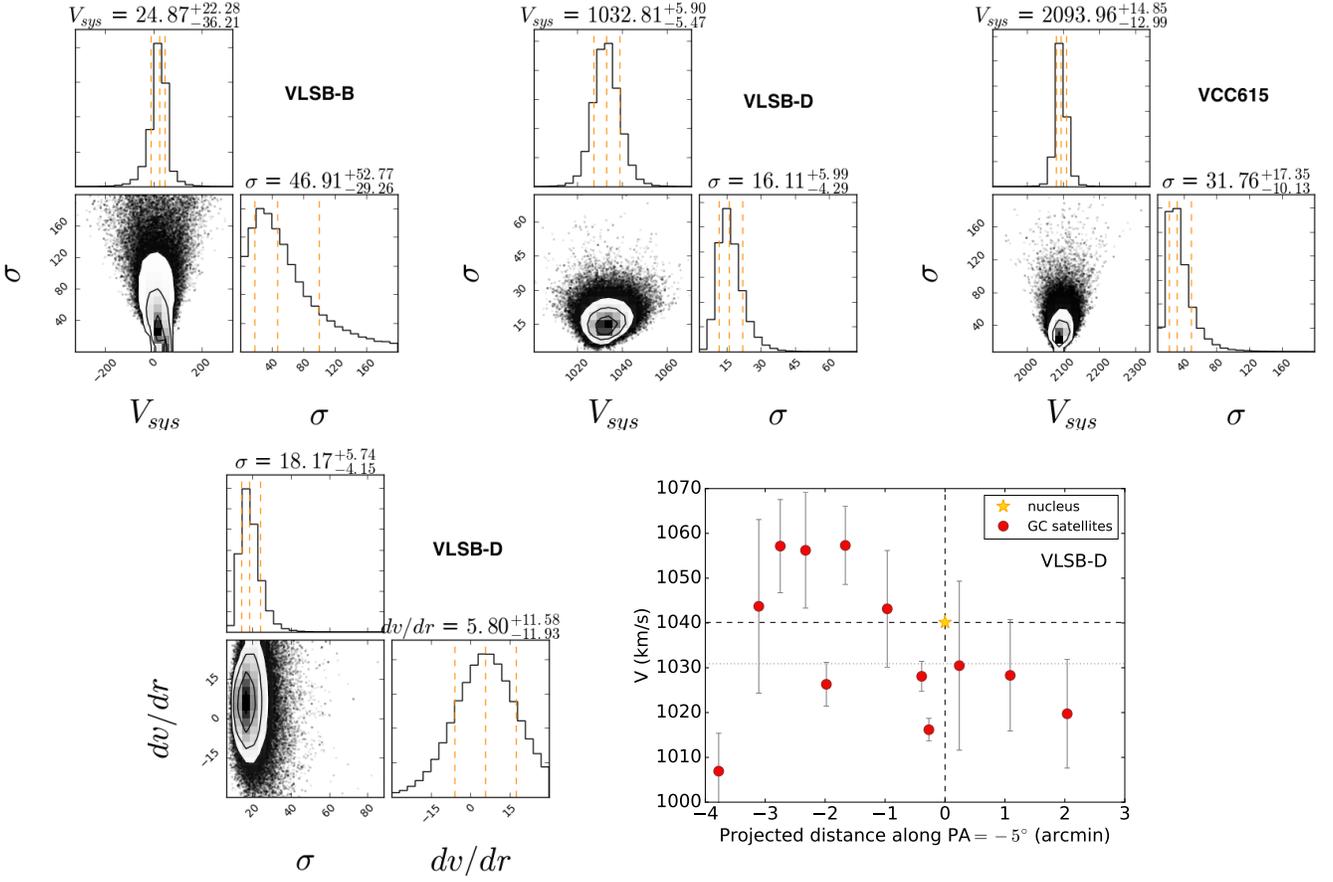}
\caption{Upper and middle panels: two-dimensional and marginalized posterior probability
  density functions for the systemic velocity ($V_{sys}$), velocity
  dispersion ($\sigma$) and velocity gradient ($dv/dr$). The orange
  lines represent the 16, 50, and 84-$th$ quartiles from left to
  right. These are considered the best value and $1\sigma$
  uncertainties. The three
  galaxies have systemic velocities consistent with being in the Virgo
cluster and dispersions within the typical values for low luminosity 
galaxies \citep[$\lesssim 50$~\kms\ for $M*<10^9$~M$_{\odot}$;][]{Geha03,Toloba14}. Lower panels: velocity
gradient for \vlsbd.}\label{MCMC}
\end{figure*}

The first time we run the MCMC following Equation~\ref{eqn2}, we
include $\phi$ as a free parameter. From that analysis we find the angle that maximizes $dv/dr$ (shown in Figure~\ref{MCMC}). In the final run, we fix
$\phi$ to this suggested position angle.

In both MCMC implementations we use flat priors within plausible
physical ranges: $V_{sys}$ is within the typical values for Virgo
cluster galaxies ($-500 < V_{sys} < 3000$~\kms);  dispersions are within
$0 < \sigma < 200$~\kms; and velocity gradients are within $-30
< dv/dr <30$~\kms~arcmin$^{-1}$

The upper panels of Figure~\ref{MCMC} show the MCMC results for Equation~\ref{eqn1}. All three
UDGs have $V_{sys}$ consistent with being galaxies in the Virgo
cluster and show a wide range of low velocity dispersions ($<
50$~\kms). The $V_{sys}$ and location in the sky suggest that \vlsbb\ and
VCC615 are members of the Virgo subcluster A within $\sim 1.1\sigma$
of its velocity distribution \citep[see][for a description of the
spectrophotometric parameters of Virgo substructures]{Boselli14}. The $V_{sys}$ of \vlsbd\  is smaller than the value
measured for its nucleus (see Figure~\ref{images} and
Table~\ref{table}). This suggests that the nucleus is not at the
center of the gravitational potential which is also supported by this nucleus being
$\sim 1$~kpc spatially off-centered. The $V_{sys}$ of VCC615 coincides with the
velocity and position of its nucleus, which suggests that it is at the center of
the gravitational potential.
 The systemic velocity of \vlsbb\  is consistent with zero,
which makes the available $Ks$
band photometry for these sources essential \citep[see
Figure~\ref{images} and][]{Munoz14}. Our four \vlsbb\ GC satellites have probabilities
of being GCs $>86 \%$ while their probabilities of being Milky Way (MW)
stars are $<6 \%$.

We perform two sets of simulations to investigate the effect that having one MW star in our sample
  of four GC satellites in \vlsbb\ would have in our measured velocity
dispersion. In the first set of simulations, we randomly select samples of three and four
objects within a Gaussian distribution with widths from 10 to
100~\kms\ in steps of 10~\kms. For each randomly selected sample we
calculate the velocity dispersion following
Equation~\ref{eqn1}. The velocity dispersions always agree within the error bars, although
the uncertainties for calculations done with three
objects are $17\%$ larger. In the second set of simulations we select
three objects from a Gaussian distribution with a width of 45~\kms,
which represent GC satellites, and one object from a Gaussian
distribution with width 100~\kms, assuming that the halo of the MW has
the same dispersion of that of M31 \citep{Gilbert14}. We also include
that the probability of this object being a MW star is $<6\%$ as
obtained from their photometric information. We calculate the
velocity dispersion of the four objects following
Equation~\ref{eqn1}. The results of these simulations suggest that we
can reject with $90\%$ confidence the hypothesis of having a MW star
in our sample. In summary, all these simulations indicate that the
probability of having a MW star in our sample of GC satellites is very
low but, if it is there, it does not affect the measured velocity
dispersion, only increases its uncertainty.

The lower panels of Figure~\ref{MCMC} show the measured velocity
gradient for \vlsbd. We use our simulations presented in
  \citet{Toloba16} to address the reliability of this velocity
  gradient given the low number statistics. These simulations show that for samples smaller than
10 GCs and velocity uncertainties $\delta v \gtrsim 10$~\kms~ (or
  $15-30\%$ relative velocity uncertainties for low-mass
  galaxies with $V_{rot}/\sigma=0--2$), the
velocity gradient measured for a galaxy that is not rotating and for a
galaxy rotating with $V_{rot}/\sigma \sim 1$ is
undistinguishable. This means that any rotation measured under these
conditions can be purely by chance. These conditions are met for
\vlsbb\ and VCC615. However, if the number of GCs is $>10$, the
average velocity uncertainty is $\delta v<10$~\kms, and the galaxy is
rotating with $V_{rot}/\sigma \sim 2$, the recovered dispersion and
rotation coincide with the input values within the error bars.
This suggests that \vlsbd\ could be rotating along its major axis,
however, due to our sample consisting only of 12 GC satellites, more
data are needed to confirm this result.

\subsection{Total Mass and Dark Matter Content}

We derive the total mass of the UDGs using the estimator for
dynamically hot systems in equilibrium by \citet{Wolf10}:

\begin{equation}
M_{1/2} = 930 \frac{\sigma^2}{km^2 s^{-2}} \frac{R_h}{pc}~M_{\odot}
\end{equation}

\noindent $\sigma$ is the velocity dispersion measured from
Equation~\ref{eqn1}, and $R_h$ is the radius that contains half the
population of the dynamical tracers. In this case, it is the radius
that contains half the number of the GCs (see Table~\ref{table}). The
diffuse nature
of the UDGs makes it challenging to decide where the GC
population ends, as a result $R_h$ is very
uncertain and it is usually assumed that $R_h=R_e$
\citep[e.g.,][]{Beasley16}. Using NGVS images, we use MCMC to fit the GC number
density profile with a Sersic function with index $n=1$ assuming
circular GC distribution. We obtain $R_h=0.85^{+0.26}_{-0.21}$~arcmin for VCC1287, which is $1.12R_e$
(assuming $R_e=45.5''$). Estimating $M_{1/2}$ using $R_h$ results in a slightly larger stellar
mass and mass-to-light ratio than that obtained by \citet{Beasley16},
$M_{1/2}=4.1^{+4.2}_{-2.7}\times10^9$~M$_{\odot}$ and $M/L_g=179^{+182}_{-117}$. For our UDGs, $R_h<R_e$ (see
Table~\ref{table}), although they are consistent within the
uncertainties. We use $R_h$ in our calculations, but if we used
$R_e$ instead, the derived $M_{1/2}$ would be $\sim 50-60\%$ larger.

\begin{figure*}
\centering
\includegraphics[angle=0,width=8.5cm]{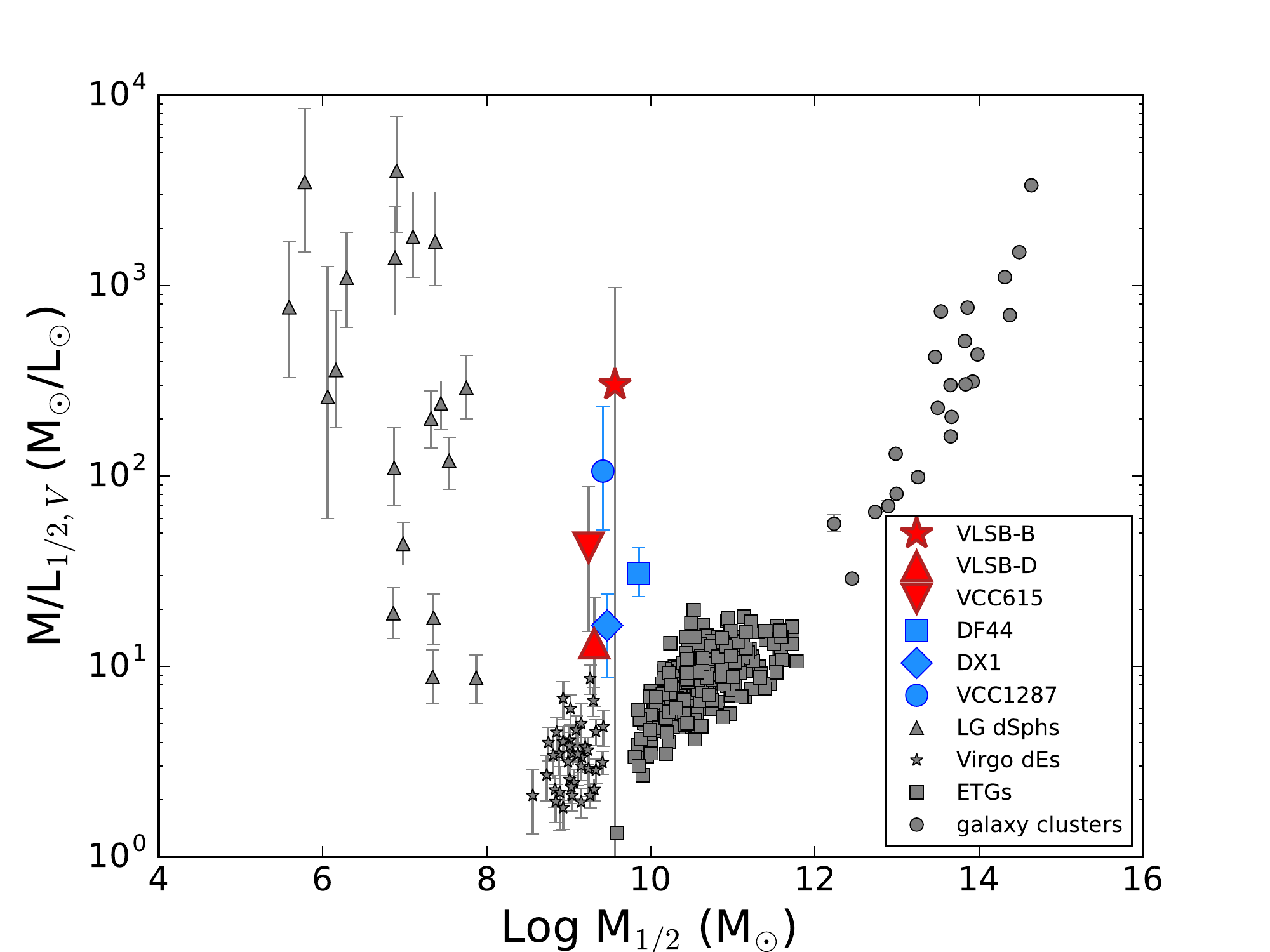}
\includegraphics[angle=0,width=8.5cm]{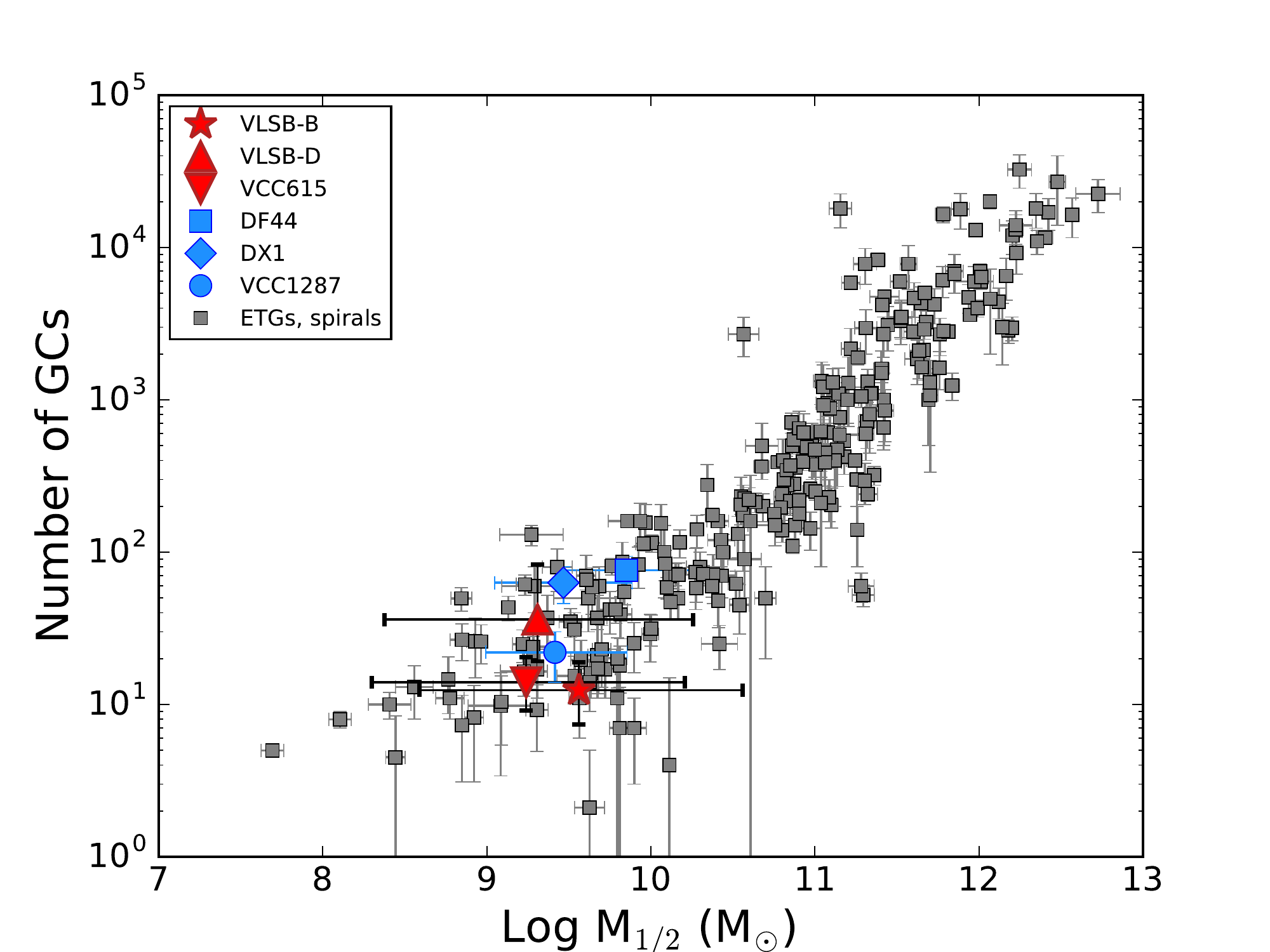}
\caption{Relations with the dynamical mass. Left panel: Mass-to-light
  ratio versus the total mass within the $R_h$. Our data is shown with
red symbols. The blue square and diamond are Dragonfly~44 and DX1 from \citet{vD17}. The
blue dot is VCC1287 from \citet{Beasley16}, this value is calculated
using the half-light radius of the galaxy instead of $R_h$. VCC1287
has $R_e>R_h$, thus, a more fair
comparison would imply moving the blue dot downwards. In gray we show the sample for
Local Group dwarf spheroidals \citep[LG dSphs][]{Wolf10}, Virgo
cluster dwarf early-types \citep[dEs][]{Toloba14}, ATLAS$^{\rm 3D}$
early-type galaxies \citep{Cappellari13}, and clusters of galaxies
\citep{Zaritsky06}. Right panel: Total number of GCs versus the total
mass within the $R_h$. Comparison sample by \citet{Jordan09,Harris13}.
}\label{Uplot}
\end{figure*}

The total masses found for the three UDGs are much higher than the
expected values for their stellar masses (see Figure~\ref{Uplot}). However, their
$N_{GC}$ are consistent with the number expected for galaxies with
that $M_{1/2}$, although \vlsbb\  appears to be on the low side of the relation.

We estimate the fraction of dark matter within the $R_h$ assuming that
these galaxies do not have gas. We use $g-i$ to estimate the
total stellar mass \citep{Taylor11} and assume that within the $R_h$
the stellar mass is half, although $R_h<R_e$, which makes the stellar
mass within the $R_h$ less than half. The results suggest these UDGs are
heavily dark matter dominated (see Table~\ref{table}).

\section{Discussion and Conclusions}\label{discussion}

\begin{figure*}
\centering
\includegraphics[angle=0,width=8.5cm]{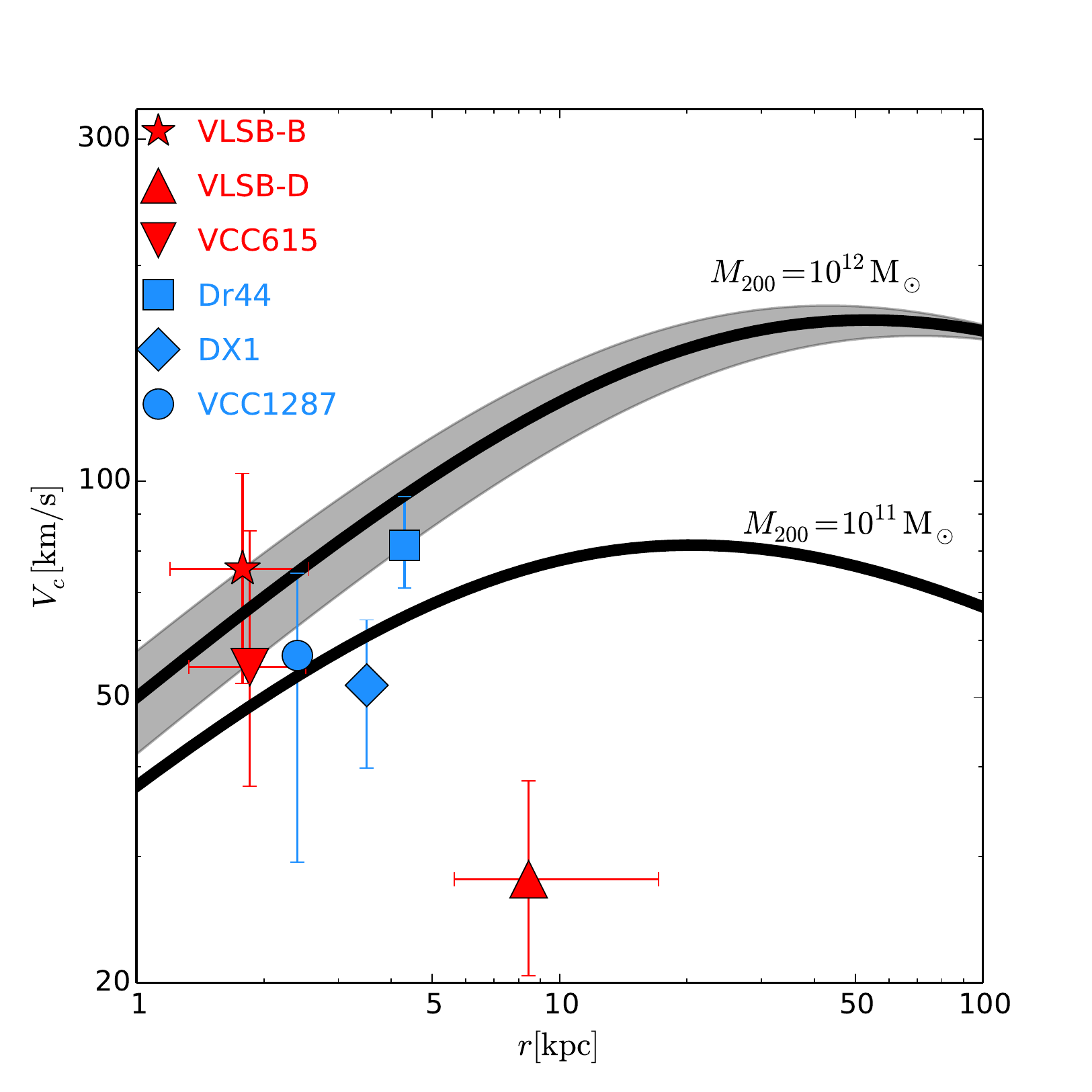}
\includegraphics[angle=0,width=8.5cm]{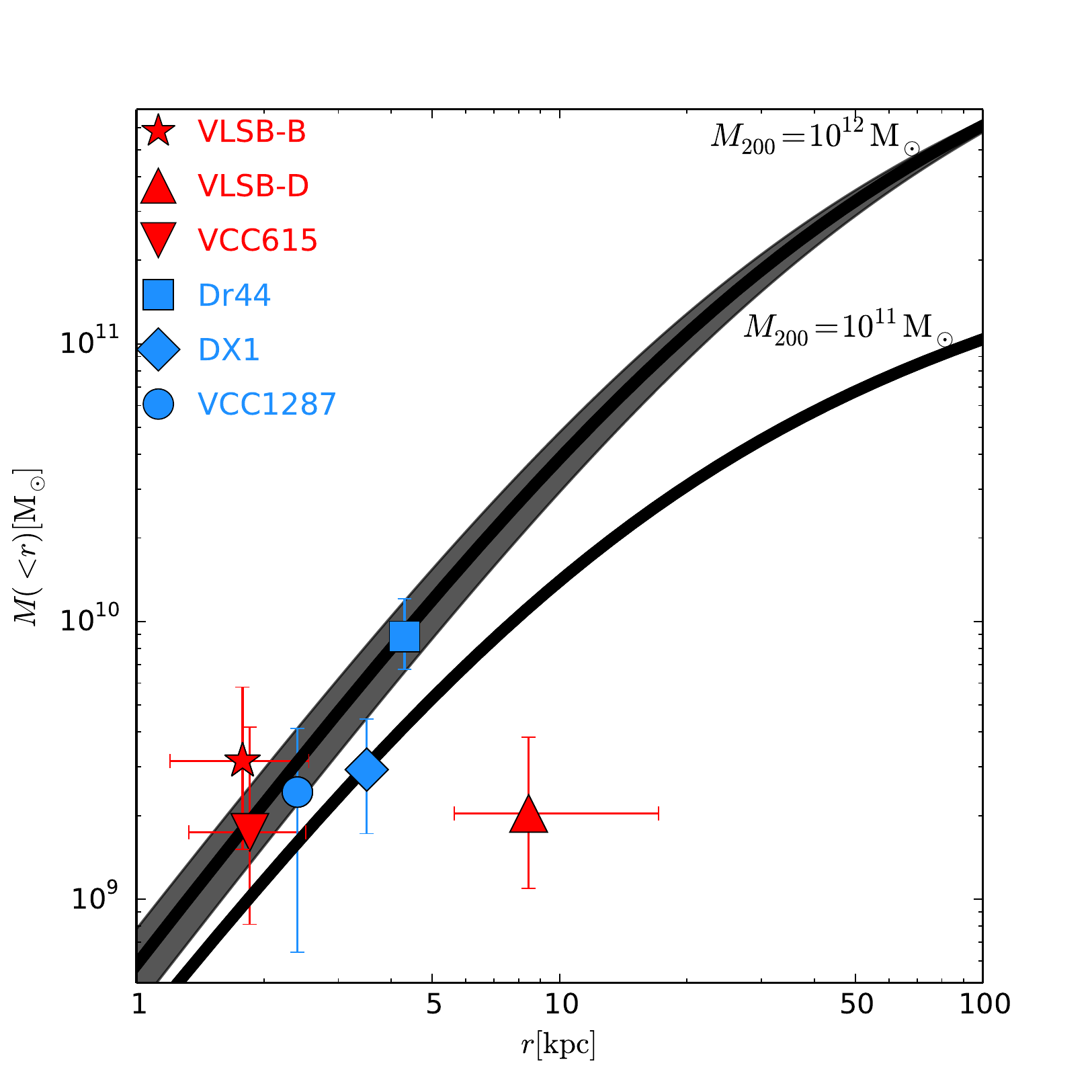}
\caption{Circular velocity (left) and total mass (right) estimates at the
  half light radius, symbols as in Figure~\ref{Uplot}. The black lines
show velocity and mass profiles corresponding to NFW halos with
fiducial masses $M_{200}=10^{11}$ and $10^{12} \; \rm M_\odot$. We
choose the average concentration $c=8.3,10$ respectively following
\citet{DuttonMaccio14}. The shaded region indicates the scatter
expected by allowing the concentration to change by $25\%$. Whereas
\vlsbd\ seems to have a lower dark matter content than expected given
its stellar mass, \vlsbb\ and VCC615 are consistent with a relatively
massive MW-like halo. }\label{DMhalo}
\end{figure*}

We spectroscopically confirm 4 GC satellites in \vlsbb, 12 in \vlsbd,
and 7 in VCC615. We use them to measure $V_{sys}$ and $\sigma$ of the
three UDGs and confirm their dynamical association
with the Virgo cluster. We estimate their total $M/L$ within the $R_h$
and find that these galaxies have extremely large values for their stellar
mass. Assuming that they follow an NFW profile \citep{NFW} where the stellar mass
is negligible as suggested by their high $M/L$, we find that \vlsbb\ and
VCC615 very likely have dark matter halos of
$\sim 10^{12}$~M$_{\odot}$ (Figure~\ref{DMhalo}). These are typical values for
galaxies that have stellar masses two orders of magnitude higher than
that of these UDGs.

The interpretation of the dark matter halo of \vlsbd\ is uncertain
given that it may not be in equilibrium. The tidal features, the
spatially and dynamically off-center nucleus, and the 
velocity gradient suggest that \vlsbd\ is being tidally stripped as it
orbits through Virgo. \vlsbd\ could have recently
interacted with M84, given their similar $V_{sys}$
\citep[1017~\kms,][]{Cappellari11}, and the fact that \vlsbd's tidal
tails align along the direction of M84.

\vlsbb\ and VCC615 show smooth and round stellar distributions
\citep{Mihos15}. If they are in dynamical equilibrium, these could be within the most
dark matter dominated galaxies known, only comparable to other UDGs
and Local Group dSphs. However, more GCs should be observed to confirm
the estimated $\sigma$. Such high $M/L$ (Figure~\ref{Uplot}) can only
be explained with massive halos and relatively high
concentrations, at least for \vlsbb\ (Figure~\ref{DMhalo}), which
might suggest an early collapse and early
infall into the cluster \citep{NFW}. This
scenario suggests that \vlsbb\ and VCC615 could be ``failed'' galaxies that
formed less stars than expected for their likely massive dark matter
halos. This could be due to an extremely low star formation efficiency
or an abrupt truncation of their star formation due to the early interaction
with the hot intracluster medium.

Our data suggests a structurally
and dynamically diverse population of galaxies, where round and
extremely low surface brightness galaxies could be rapidly
rotating. It is important to statistically quantify the significance of
such rotation not only in \vlsbb\ but also in other UDGs by 
increasing the number of GCs observed and the
number of UDGs studied dynamically.

\acknowledgments

E.T acknowledges the support from the Eberhardt Fellowship awarded by
the University of the Pacific. E.T. and P.G. acknowledge the NSF grants AST-1010039 and AST-1412504.  SL and EWP acknowledge support from
NSFC grant 11573002. LVS acknowledges support from HST-AR-14583 and
the Hellman Foundation. The authors thank the referee for useful suggestions that have helped to improve this manuscript.

\bibliographystyle{aa}
\bibliography{references}{}

\begin{thebibliography}{51}
\expandafter\ifx\csname natexlab\endcsname\relax\def\natexlab#1{#1}\fi

\bibitem[{{Amorisco} \& {Loeb}(2016)}]{AmoriscoLoeb16}
{Amorisco}, N.~C. \& {Loeb}, A. 2016, \mnras, 459, L51

\bibitem[{{Amorisco} {et~al.}(2016){Amorisco}, {Monachesi}, \&
  {White}}]{Amorisco17}
{Amorisco}, N.~C., {Monachesi}, A., \& {White}, S.~D.~M. 2016, ArXiv e-prints

\bibitem[{{Beasley} {et~al.}(2016){Beasley}, {Romanowsky}, {Pota}, {Navarro},
  {Martinez Delgado}, {Neyer}, \& {Deich}}]{Beasley16}
{Beasley}, M.~A., {Romanowsky}, A.~J., {Pota}, V., {et~al.} 2016, \apjl, 819,
  L20

\bibitem[{{Binggeli} {et~al.}(1987){Binggeli}, {Tammann}, \&
  {Sandage}}]{Binggeli87}
{Binggeli}, B., {Tammann}, G.~A., \& {Sandage}, A. 1987, \aj, 94, 251

\bibitem[{{Blakeslee} {et~al.}(2009){Blakeslee}, {Jord{\'a}n}, {Mei},
  {C{\^o}t{\'e}}, {Ferrarese}, {Infante}, {Peng}, {Tonry}, \&
  {West}}]{Blakeslee09}
{Blakeslee}, J.~P., {Jord{\'a}n}, A., {Mei}, S., {et~al.} 2009, \apj, 694, 556

\bibitem[{{Boselli} {et~al.}(2014){Boselli}, {Voyer}, {Boissier}, {Cucciati},
  {Consolandi}, {Cortese}, {Fumagalli}, {Gavazzi}, {Heinis}, {Roehlly}, \&
  {Toloba}}]{Boselli14}
{Boselli}, A., {Voyer}, E., {Boissier}, S., {et~al.} 2014, \aap, 570, A69

\bibitem[{{Caldwell}(2006)}]{Caldwell06}
{Caldwell}, N. 2006, \apj, 651, 822

\bibitem[{{Cappellari} \& {Emsellem}(2004)}]{ppxf}
{Cappellari}, M. \& {Emsellem}, E. 2004, \pasp, 116, 138

\bibitem[{{Cappellari} {et~al.}(2011){Cappellari}, {Emsellem}, {Krajnovi{\'c}},
  {McDermid}, {Scott}, {Verdoes Kleijn}, {Young}, {Alatalo}, {Bacon}, {Blitz},
  {Bois}, {Bournaud}, {Bureau}, {Davies}, {Davis}, {de Zeeuw}, {Duc},
  {Khochfar}, {Kuntschner}, {Lablanche}, {Morganti}, {Naab}, {Oosterloo},
  {Sarzi}, {Serra}, \& {Weijmans}}]{Cappellari11}
{Cappellari}, M., {Emsellem}, E., {Krajnovi{\'c}}, D., {et~al.} 2011, \mnras,
  413, 813

\bibitem[{{Cappellari} {et~al.}(2013){Cappellari}, {McDermid}, {Alatalo},
  {Blitz}, {Bois}, {Bournaud}, {Bureau}, {Crocker}, {Davies}, {Davis}, {de
  Zeeuw}, {Duc}, {Emsellem}, {Khochfar}, {Krajnovi{\'c}}, {Kuntschner},
  {Morganti}, {Naab}, {Oosterloo}, {Sarzi}, {Scott}, {Serra}, {Weijmans}, \&
  {Young}}]{Cappellari13}
{Cappellari}, M., {McDermid}, R.~M., {Alatalo}, K., {et~al.} 2013, \mnras, 432,
  1862

\bibitem[{{Cooper} {et~al.}(2012){Cooper}, {Newman}, {Davis}, {Finkbeiner}, \&
  {Gerke}}]{DEIMOSpipeline1}
{Cooper}, M.~C., {Newman}, J.~A., {Davis}, M., {Finkbeiner}, D.~P., \& {Gerke},
  B.~F. 2012, {spec2d: DEEP2 DEIMOS Spectral Pipeline}, astrophysics Source
  Code Library, ascl:1203.003

\bibitem[{{Dalcanton} {et~al.}(1997){Dalcanton}, {Spergel}, {Gunn}, {Schmidt},
  \& {Schneider}}]{Dalcanton97}
{Dalcanton}, J.~J., {Spergel}, D.~N., {Gunn}, J.~E., {Schmidt}, M., \&
  {Schneider}, D.~P. 1997, \aj, 114, 635

\bibitem[{{Di Cintio} {et~al.}(2017){Di Cintio}, {Brook}, {Dutton},
  {Macci{\`o}}, {Obreja}, \& {Dekel}}]{DiCintio17}
{Di Cintio}, A., {Brook}, C.~B., {Dutton}, A.~A., {et~al.} 2017, \mnras, 466,
  L1

\bibitem[{{Dutton} \& {Macci{\`o}}(2014)}]{DuttonMaccio14}
{Dutton}, A.~A. \& {Macci{\`o}}, A.~V. 2014, \mnras, 441, 3359

\bibitem[{{Faber} {et~al.}(2003){Faber}, {Phillips}, {Kibrick}, {Alcott},
  {Allen}, {Burrous}, {Cantrall}, {Clarke}, {Coil}, {Cowley}, {Davis}, {Deich},
  {Dietsch}, {Gilmore}, {Harper}, {Hilyard}, {Lewis}, {McVeigh}, {Newman},
  {Osborne}, {Schiavon}, {Stover}, {Tucker}, {Wallace}, {Wei}, {Wirth}, \&
  {Wright}}]{Faber03}
{Faber}, S.~M., {Phillips}, A.~C., {Kibrick}, R.~I., {et~al.} 2003, in
  \procspie, Vol. 4841, Instrument Design and Performance for Optical/Infrared
  Ground-based Telescopes, ed. M.~{Iye} \& A.~F.~M. {Moorwood}, 1657--1669

\bibitem[{{Ferrarese} {et~al.}(2012){Ferrarese}, {C{\^o}t{\'e}}, {Cuillandre},
  {Gwyn}, {Peng}, {MacArthur}, {Duc}, {Boselli}, {Mei}, {Erben}, {McConnachie},
  {Durrell}, {Mihos}, {Jord{\'a}n}, {Lan{\c c}on}, {Puzia}, {Emsellem},
  {Balogh}, {Blakeslee}, {van Waerbeke}, {Gavazzi}, {Vollmer}, {Kavelaars},
  {Woods}, {Ball}, {Boissier}, {Courteau}, {Ferriere}, {Gavazzi},
  {Hildebrandt}, {Hudelot}, {Huertas-Company}, {Liu}, {McLaughlin}, {Mellier},
  {Milkeraitis}, {Schade}, {Balkowski}, {Bournaud}, {Carlberg}, {Chapman},
  {Hoekstra}, {Peng}, {Sawicki}, {Simard}, {Taylor}, {Tully}, {van Driel},
  {Wilson}, {Burdullis}, {Mahoney}, \& {Manset}}]{Ferrarese12}
{Ferrarese}, L., {C{\^o}t{\'e}}, P., {Cuillandre}, J.-C., {et~al.} 2012, \apjs,
  200, 4

\bibitem[{{Foreman-Mackey} {et~al.}(2013){Foreman-Mackey}, {Hogg}, {Lang}, \&
  {Goodman}}]{emcee}
{Foreman-Mackey}, D., {Hogg}, D.~W., {Lang}, D., \& {Goodman}, J. 2013, \pasp,
  125, 306

\bibitem[{{Geha} {et~al.}(2003){Geha}, {Guhathakurta}, \& {van der
  Marel}}]{Geha03}
{Geha}, M., {Guhathakurta}, P., \& {van der Marel}, R.~P. 2003, \aj, 126, 1794

\bibitem[{{Gilbert} {et~al.}(2014){Gilbert}, {Kalirai}, {Guhathakurta},
  {Beaton}, {Geha}, {Kirby}, {Majewski}, {Patterson}, {Tollerud}, {Bullock},
  {Tanaka}, \& {Chiba}}]{Gilbert14}
{Gilbert}, K.~M., {Kalirai}, J.~S., {Guhathakurta}, P., {et~al.} 2014, \apj,
  796, 76

\bibitem[{{Harris} {et~al.}(2013){Harris}, {Harris}, \& {Alessi}}]{Harris13}
{Harris}, W.~E., {Harris}, G.~L.~H., \& {Alessi}, M. 2013, \apj, 772, 82

\bibitem[{{Impey} {et~al.}(1988){Impey}, {Bothun}, \& {Malin}}]{Impey88}
{Impey}, C., {Bothun}, G., \& {Malin}, D. 1988, \apj, 330, 634

\bibitem[{{Janssens} {et~al.}(2017){Janssens}, {Abraham}, {Brodie}, {Forbes},
  {Romanowsky}, \& {van Dokkum}}]{Janssens17}
{Janssens}, S., {Abraham}, R., {Brodie}, J., {et~al.} 2017, \apjl, 839, L17

\bibitem[{{Jord{\'a}n} {et~al.}(2009){Jord{\'a}n}, {Peng}, {Blakeslee},
  {C{\^o}t{\'e}}, {Eyheramendy}, {Ferrarese}, {Mei}, {Tonry}, \&
  {West}}]{Jordan09}
{Jord{\'a}n}, A., {Peng}, E.~W., {Blakeslee}, J.~P., {et~al.} 2009, \apjs, 180,
  54

\bibitem[{{Kirby} {et~al.}(2015{\natexlab{a}}){Kirby}, {Guo}, {Zhang}, {Deng},
  {Cohen}, {Guhathakurta}, {Shetrone}, {Lee}, \& {Rizzi}}]{Kirby15a}
{Kirby}, E.~N., {Guo}, M., {Zhang}, A.~J., {et~al.} 2015{\natexlab{a}}, \apj,
  801, 125

\bibitem[{{Kirby} {et~al.}(2015{\natexlab{b}}){Kirby}, {Simon}, \&
  {Cohen}}]{Kirby15b}
{Kirby}, E.~N., {Simon}, J.~D., \& {Cohen}, J.~G. 2015{\natexlab{b}}, \apj,
  810, 56

\bibitem[{{Koda} {et~al.}(2015){Koda}, {Yagi}, {Yamanoi}, \&
  {Komiyama}}]{Koda15}
{Koda}, J., {Yagi}, M., {Yamanoi}, H., \& {Komiyama}, Y. 2015, \apjl, 807, L2

\bibitem[{{Martin} \& {Jin}(2010)}]{Martin10}
{Martin}, N.~F. \& {Jin}, S. 2010, \apj, 721, 1333

\bibitem[{{Mart{\'{\i}}nez-Delgado} {et~al.}(2016){Mart{\'{\i}}nez-Delgado},
  {L{\"a}sker}, {Sharina}, {Toloba}, {Fliri}, {Beaton}, {Valls-Gabaud},
  {Karachentsev}, {Chonis}, {Grebel}, {Forbes}, {Romanowsky},
  {Gallego-Laborda}, {Teuwen}, {G{\'o}mez-Flechoso}, {Wang}, {Guhathakurta},
  {Kaisin}, \& {Ho}}]{MartinezDelgado16}
{Mart{\'{\i}}nez-Delgado}, D., {L{\"a}sker}, R., {Sharina}, M., {et~al.} 2016,
  \aj, 151, 96

\bibitem[{{Mei} {et~al.}(2007){Mei}, {Blakeslee}, {C{\^o}t{\'e}}, {Tonry},
  {West}, {Ferrarese}, {Jord{\'a}n}, {Peng}, {Anthony}, \& {Merritt}}]{Mei07}
{Mei}, S., {Blakeslee}, J.~P., {C{\^o}t{\'e}}, P., {et~al.} 2007, \apj, 655,
  144

\bibitem[{{Mihos} {et~al.}(2015){Mihos}, {Durrell}, {Ferrarese}, {Feldmeier},
  {C{\^o}t{\'e}}, {Peng}, {Harding}, {Liu}, {Gwyn}, \& {Cuillandre}}]{Mihos15}
{Mihos}, J.~C., {Durrell}, P.~R., {Ferrarese}, L., {et~al.} 2015, \apjl, 809,
  L21

\bibitem[{{Mihos} {et~al.}(2017){Mihos}, {Harding}, {Feldmeier}, {Rudick},
  {Janowiecki}, {Morrison}, {Slater}, \& {Watkins}}]{Mihos17}
{Mihos}, J.~C., {Harding}, P., {Feldmeier}, J.~J., {et~al.} 2017, \apj, 834, 16

\bibitem[{{Moore} {et~al.}(1996){Moore}, {Katz}, {Lake}, {Dressler}, \&
  {Oemler}}]{Moore96}
{Moore}, B., {Katz}, N., {Lake}, G., {Dressler}, A., \& {Oemler}, A. 1996,
  \nat, 379, 613

\bibitem[{{Mu{\~n}oz} {et~al.}(2015){Mu{\~n}oz}, {Eigenthaler}, {Puzia},
  {Taylor}, {Ordenes-Brice{\~n}o}, {Alamo-Mart{\'{\i}}nez}, {Ribbeck},
  {{\'A}ngel}, {Capaccioli}, {C{\^o}t{\'e}}, {Ferrarese}, {Galaz}, {Hempel},
  {Hilker}, {Jord{\'a}n}, {Lan{\c c}on}, {Mieske}, {Paolillo}, {Richtler},
  {S{\'a}nchez-Janssen}, \& {Zhang}}]{Munoz15}
{Mu{\~n}oz}, R.~P., {Eigenthaler}, P., {Puzia}, T.~H., {et~al.} 2015, \apjl,
  813, L15

\bibitem[{{Mu{\~n}oz} {et~al.}(2014){Mu{\~n}oz}, {Puzia}, {Lan{\c c}on},
  {Peng}, {C{\^o}t{\'e}}, {Ferrarese}, {Blakeslee}, {Mei}, {Cuillandre},
  {Hudelot}, {Courteau}, {Duc}, {Balogh}, {Boselli}, {Bournaud}, {Carlberg},
  {Chapman}, {Durrell}, {Eigenthaler}, {Emsellem}, {Gavazzi}, {Gwyn},
  {Huertas-Company}, {Ilbert}, {Jord{\'a}n}, {L{\"a}sker}, {Licitra}, {Liu},
  {MacArthur}, {McConnachie}, {McCracken}, {Mellier}, {Peng}, {Raichoor},
  {Taylor}, {Tonry}, {Tully}, \& {Zhang}}]{Munoz14}
{Mu{\~n}oz}, R.~P., {Puzia}, T.~H., {Lan{\c c}on}, A., {et~al.} 2014, \apjs,
  210, 4

\bibitem[{{Navarro} {et~al.}(1997){Navarro}, {Frenk}, \& {White}}]{NFW}
{Navarro}, J.~F., {Frenk}, C.~S., \& {White}, S.~D.~M. 1997, \apj, 490, 493

\bibitem[{{Newman} {et~al.}(2013){Newman}, {Cooper}, {Davis}, {Faber}, {Coil},
  {Guhathakurta}, {Koo}, {Phillips}, {Conroy}, {Dutton}, {Finkbeiner}, {Gerke},
  {Rosario}, {Weiner}, {Willmer}, {Yan}, {Harker}, {Kassin}, {Konidaris},
  {Lai}, {Madgwick}, {Noeske}, {Wirth}, {Connolly}, {Kaiser}, {Kirby},
  {Lemaux}, {Lin}, {Lotz}, {Luppino}, {Marinoni}, {Matthews}, {Metevier}, \&
  {Schiavon}}]{DEIMOSpipeline2}
{Newman}, J.~A., {Cooper}, M.~C., {Davis}, M., {et~al.} 2013, \apjs, 208, 5

\bibitem[{{Peng} {et~al.}(2006){Peng}, {Jord{\'a}n}, {C{\^o}t{\'e}},
  {Blakeslee}, {Ferrarese}, {Mei}, {West}, {Merritt}, {Milosavljevi{\'c}}, \&
  {Tonry}}]{Peng06}
{Peng}, E.~W., {Jord{\'a}n}, A., {C{\^o}t{\'e}}, P., {et~al.} 2006, \apj, 639,
  95

\bibitem[{{Powalka} {et~al.}(2016){Powalka}, {Lan{\c c}on}, {Puzia}, {Peng},
  {Liu}, {Mu{\~n}oz}, {Blakeslee}, {C{\^o}t{\'e}}, {Ferrarese}, {Roediger},
  {S{\'a}nchez-Janssen}, {Zhang}, {Durrell}, {Cuillandre}, {Duc},
  {Guhathakurta}, {Gwyn}, {Hudelot}, {Mei}, \& {Toloba}}]{Powalka16}
{Powalka}, M., {Lan{\c c}on}, A., {Puzia}, T.~H., {et~al.} 2016, \apjs, 227, 12

\bibitem[{{Rom{\'a}n} \& {Trujillo}(2017)}]{RomanTrujillo17}
{Rom{\'a}n}, J. \& {Trujillo}, I. 2017, \mnras, 468, 703

\bibitem[{{Sandage} \& {Binggeli}(1984)}]{SandageBinggeli84}
{Sandage}, A. \& {Binggeli}, B. 1984, \aj, 89, 919

\bibitem[{{Taylor} {et~al.}(2011){Taylor}, {Hopkins}, {Baldry}, {Brown},
  {Driver}, {Kelvin}, {Hill}, {Robotham}, {Bland-Hawthorn}, {Jones}, {Sharp},
  {Thomas}, {Liske}, {Loveday}, {Norberg}, {Peacock}, {Bamford}, {Brough},
  {Colless}, {Cameron}, {Conselice}, {Croom}, {Frenk}, {Gunawardhana},
  {Kuijken}, {Nichol}, {Parkinson}, {Phillipps}, {Pimbblet}, {Popescu},
  {Prescott}, {Sutherland}, {Tuffs}, {van Kampen}, \& {Wijesinghe}}]{Taylor11}
{Taylor}, E.~N., {Hopkins}, A.~M., {Baldry}, I.~K., {et~al.} 2011, \mnras, 418,
  1587

\bibitem[{{Toloba} {et~al.}(2014){Toloba}, {Guhathakurta}, {Peletier},
  {Boselli}, {Lisker}, {Falc{\'o}n-Barroso}, {Simon}, {van de Ven}, {Paudel},
  {Emsellem}, {Janz}, {den Brok}, {Gorgas}, {Hensler}, {Laurikainen}, {Niemi},
  {Ry{\'s}}, \& {Salo}}]{Toloba14}
{Toloba}, E., {Guhathakurta}, P., {Peletier}, R.~F., {et~al.} 2014, \apjs, 215,
  17

\bibitem[{{Toloba} {et~al.}(2016{\natexlab{a}}){Toloba}, {Li}, {Guhathakurta},
  {Peng}, {Ferrarese}, {C{\^o}t{\'e}}, {Emsellem}, {Gwyn}, {Zhang}, {Boselli},
  {Cuillandre}, {Jordan}, \& {Liu}}]{Toloba16}
{Toloba}, E., {Li}, B., {Guhathakurta}, P., {et~al.} 2016{\natexlab{a}}, \apj,
  822, 51

\bibitem[{{Toloba} {et~al.}(2016{\natexlab{b}}){Toloba}, {Sand}, {Spekkens},
  {Crnojevi{\'c}}, {Simon}, {Guhathakurta}, {Strader}, {Caldwell}, {McLeod}, \&
  {Seth}}]{Toloba16b}
{Toloba}, E., {Sand}, D.~J., {Spekkens}, K., {et~al.} 2016{\natexlab{b}},
  \apjl, 816, L5

\bibitem[{{van der Burg} {et~al.}(2016){van der Burg}, {Muzzin}, \&
  {Hoekstra}}]{vanderburg16}
{van der Burg}, R.~F.~J., {Muzzin}, A., \& {Hoekstra}, H. 2016, \aap, 590, A20

\bibitem[{{van Dokkum} {et~al.}(2016){van Dokkum}, {Abraham}, {Brodie},
  {Conroy}, {Danieli}, {Merritt}, {Mowla}, {Romanowsky}, \& {Zhang}}]{vD16}
{van Dokkum}, P., {Abraham}, R., {Brodie}, J., {et~al.} 2016, \apjl, 828, L6

\bibitem[{{van Dokkum} {et~al.}(2017){van Dokkum}, {Abraham}, {Romanowsky},
  {Brodie}, {Conroy}, {Danieli}, {Lokhorst}, {Merritt}, {Mowla}, \&
  {Zhang}}]{vD17}
{van Dokkum}, P., {Abraham}, R., {Romanowsky}, A.~J., {et~al.} 2017, \apjl,
  844, L11

\bibitem[{{van Dokkum} {et~al.}(2015){van Dokkum}, {Abraham}, {Merritt},
  {Zhang}, {Geha}, \& {Conroy}}]{vD15a}
{van Dokkum}, P.~G., {Abraham}, R., {Merritt}, A., {et~al.} 2015, \apjl, 798,
  L45

\bibitem[{{Venhola} {et~al.}(2017){Venhola}, {Peletier}, {Laurikainen}, {Salo},
  {Lisker}, {Iodice}, {Capaccioli}, {Kleijn}, {Valentijn}, {Mieske}, {Hilker},
  {Wittmann}, {van de Ven}, {Grado}, {Spavone}, {Cantiello}, {Napolitano},
  {Paolillo}, \& {Falc{\'o}n-Barroso}}]{venhola17}
{Venhola}, A., {Peletier}, R., {Laurikainen}, E., {et~al.} 2017, \aap, 608,
  A142

\bibitem[{{Wolf} {et~al.}(2010){Wolf}, {Martinez}, {Bullock}, {Kaplinghat},
  {Geha}, {Mu{\~n}oz}, {Simon}, \& {Avedo}}]{Wolf10}
{Wolf}, J., {Martinez}, G.~D., {Bullock}, J.~S., {et~al.} 2010, \mnras, 406,
  1220

\bibitem[{{Zaritsky} {et~al.}(2006){Zaritsky}, {Gonzalez}, \&
  {Zabludoff}}]{Zaritsky06}
{Zaritsky}, D., {Gonzalez}, A.~H., \& {Zabludoff}, A.~I. 2006, \apj, 638, 725

\end{thebibliography}


\end{document}